\documentclass[11pt,a4paper]{article} % JCAP format
\usepackage{jcappub}

\usepackage{graphicx}

\title{Curvature perturbation spectra from waterfall transition, black
hole constraints and non-Gaussianity}
\author{Edgar Bugaev}
\author{and Peter Klimai}
\affiliation{Institute for Nuclear Research, Russian Academy of
Sciences, 60th October Anniversary Prospect 7a, 117312 Moscow, Russia}

\emailAdd{bugaev@pcbai10.inr.ruhep.ru}
\emailAdd{pklimai@gmail.com}

\abstract{
We carried out numerical calculations of a contribution of the
waterfall field to the primordial curvature perturbation (on uniform
density hypersurfaces) $\zeta$, which is produced during waterfall
transition in hybrid inflation scenario. The calculation is performed
for a broad interval of values of the model parameters. We show that
there is a strong growth of amplitudes of the curvature perturbation
spectrum in the limit when the bare mass-squared of the waterfall
field becomes comparable with the square of Hubble parameter.
We show that in this limit the primordial black hole constraints
on the curvature perturbations must be taken into account.
It is shown that, in the same limit, peak values of the curvature
perturbation spectra are far beyond horizon, and the spectra are
strongly non-Gaussian.
}

\arxivnumber{1107.3754}

\begin{document}

\maketitle

\section{Introduction}
\label{sec-intro}

In last two years an interest in hybrid inflation models was very much revived
\cite{Lyth:2010ch, Gong:2010zf, Fonseca:2010nk,
Abolhasani:2010kr, Abolhasani:2010kn, Lyth:2010zq, Abolhasani:2011yp, Lyth:2011kj}
(see also earlier works \cite{Barnaby:2006cq, Barnaby:2006km}).
The main questions which were discussed are dynamics of the waterfall field
and its influence on the total spectrum of density perturbations produced by
inflationary expansion. It was suggested, in particular, that fluctuations
in the waterfall field could lead to non-Gaussian curvature perturbation
at rather large scales (even, possibly, at cosmological scales).
Note that, in general, hybrid inflation models remain theoretically attractive up to now,
especially in the context of supergravity and string theories (see, e.g.,
\cite{Kallosh:2007ig, Kallosh:2003ux}).

In many cosmological scenarios the period just after the end of inflation,
i.e., the inflaton decay and the subsequent evolution of the decay products
to a thermal equilibrium starts with preheating
\cite{Traschen:1990sw, Dolgov:1989us, Kofman:1994rk, Shtanov:1994ce}.
One of the most studied models is hybrid inflation with tachyonic preheating. It
had been shown in \cite{Felder:2000hj} that preheating after hybrid inflation
goes through the tachyonic amplification due to the dynamical symmetry
breaking, when one of the fields rolls to the minimum through the region
where its effective mass-squared is negative. In the process of this rolling
amplitudes of field fluctuations grow exponentially leading to a fast
decrease of a height of the inflationary potential (``false vacuum decay'').

As is shown in the recent papers \cite{Lyth:2010ch, Gong:2010zf, Fonseca:2010nk,
Abolhasani:2010kr, Lyth:2010zq, Abolhasani:2011yp, Lyth:2011kj} the power
spectrum of the comoving curvature perturbations from the waterfall field in
hybrid inflation with tachyonic preheating is very blue: it depends
on comoving wavenumber $k$ like $(k/k_*)^3$ (for $k<k_*$) and is negligible
on the cosmological scales. Absolute value of the spectrum amplitude
at $k\sim k_*$ (as well as the value of $k_*$) depends on model parameters.
In principle, this spectrum at $k\sim k_*$ may be quite substantial, and,
in this case, it may be constrained by data of primordial black hole (PBH)
and relict gravitational wave (GW) searches.

The amplification of curvature perturbations after preheating (e.g.,
in two-field inflation models) and blue spectra of the curvature perturbations
on the constant energy density hypersurfaces, $\zeta$, behaving like $\sim k^3$
on super-horizon scales (in a case of the quadratic inflationary potential)
had been predicted in many works (see, e.g.,
\cite{Bassett:1998wg, Liddle:1999hq, Green:2000he}). Another example where
the steeply blue ($\sim k^3$) curvature perturbation spectrum is predicted
is one of models of false vacuum inflation, in which
the inflationary potential has a local
minimum for a trapping of the field and a high temperature correction for
a termination of inflation \cite{Pilo:2004ke, Gong:2008ni}.

In more recent studies of preheating processes
\cite{Suyama:2004mz, Suyama:2006sr, Brax:2010ai, Dufaux:2008dn},
primordial density perturbation spectra with blue tilt or,
more generally, spectra having broad peak features at some $k_*$ value are predicted
in two-field models of inflation ending by preheating (going through the
parametric resonance) \cite{Suyama:2004mz} and even in one-field inflation
models (in particular, in models with tachyonic preheating after small-field
inflation \cite{Suyama:2006sr, Brax:2010ai}). The characteristic values of
$k_*$ (i.e., the $k$-values of modes which become dominant as a result of
tachyonic amplification or parametric resonance) are
different in different models. In preheating after
chaotic inflation \cite{Kofman:1994rk, Kofman:1997yn} where the perturbations are
amplified by parametric resonance, power spectra are peaked at scales $k/aH \gg 1$,
whereas in models of small-field inflation the spectra may be peaked around the Hubble
scale. In preheating after hybrid inflation, the typical scale of $k$-values
amplified by the tachyonic instability must be sub-Hubble if the phase transition
is fast, i.e., if it takes less than a Hubble time \cite{Lyth:2010ch, Dufaux:2008dn}.
Clearly, the $k$-value of the dominantly amplified mode is an important
characteristic of a preheating model because large curvature and
density perturbations of the Hubble size may result in an abundant
production of PBHs and GWs.

In the first stage of tachyonic preheating in models of hybrid inflation the
amplification of initial quantum fluctuations of the non-inflaton field
is realized due to the classical inflaton rolling or, in a case of the small initial
velocity of the inflaton field, due to processes of quantum diffusion \cite{Felder:2000hj}.
The former case is characterized by the existence of a period of the linear evolution
of the non-inflaton field. The evolution in this case can be studied using the cosmological
perturbation theory \cite{Asaka:2001ez, Copeland:2002ku, GarciaBellido:2002aj}
or $\delta N$ approach \cite{Starobinsky:1982ee, Starobinsky:1986fxa, Sasaki:1995aw, Lyth:2004gb}.

In the present paper we want to study in detail the dependence of the
primordial curvature perturbations produced by hybrid inflation with tachyonic
preheating on parameters of the inflationary potential. We show, in particular,
that perturbation amplitudes strongly depend on the mass of the waterfall
field $m_\chi$. More exactly, it depends on the relation $|m_\chi^2/H^2|$, where $H$ is
a value of the Hubble parameter during inflation. At small values of
this relation, $|m_\chi^2/H^2|\sim 1$, the waterfall transition is rather slow,
and the expansion of the Universe can not be ignored. Our main aim is to
predict concrete values of the perturbation spectrum amplitudes, as well
as a form of the $k$-dependence of the spectrum. So, we preferred to use the
results of the numerical (rather than analytical) solution of the equations for the time evolution
of the waterfall field. The approximate analytic expression for the power
spectrum amplitude, according to which one has, roughly, \cite{Lyth:2011kj}
\begin{equation}
\label{Pzeta1}
{\cal P}_{\zeta} \sim \left| \frac{H^2}{m_\chi^2} \right| \cdot \left( \frac{k}{k_*} \right)^3,
\;\;\;\;\; k<k_*,
\end{equation}
was obtained in the limit of the fast waterfall transition, when $|m_\chi^2| \gg H^2$,
and can not be a priori used in a whole region of the parameter space. This formula works well
when number of e-folds of cosmological expansion during waterfall is not too small
(say, $N \gtrsim 0.1$). If $N\ll 0.1$, eq. (\ref{Pzeta1}) must be multiplied on the factor
which is proportional to $N$ and can be much less than $1$ \cite{Lyth:2011kj}.

The plan of the paper is as follows. In section \ref{sec-wf} we calculate the
time evolution of the power spectra of the waterfall field perturbations and
the spectra at the end of the waterfall.
The curvature perturbation spectra from the waterfall, as a function
of model parameters, are calculated in section \ref{sec-zeta}.
In section \ref{sec-pbh} we give estimates for the possibility of PBH production
in hybrid inflation model, with taking into account the non-Gaussianity
of produced perturbations. Section \ref{sec-concl} contains our conclusions.

\section{ Calculation of waterfall field amplitudes }
\label{sec-wf}

We consider the hybrid inflation model which describes an evolution of the slowly rolling
inflaton field  $\phi$ and the waterfall field $\chi$, with the
potential \cite{Linde:1991km, Linde:1993cn}
\begin{equation}
\label{pot}
V(\phi, \chi)= \left( M^2 - \frac{\sqrt{\lambda}}{2} \chi^2 \right)^2 + \frac{1}{2}m^2\phi^2
+ \frac{1}{2} \gamma \phi^2 \chi^2 .
\end{equation}
The first term in eq. (\ref{pot}) is a potential for the waterfall field $\chi$ with the false
vacuum at $\chi=0$ and true vacuum at $\chi_0^2=2 M^2/\sqrt{\lambda} \equiv v^2$. The
effective mass of the waterfall field in the false vacuum state is given by
\begin{equation}
\label{phic-gamma}
m_\chi^2(\phi) = \gamma \left( \phi^2 - \phi_c^2 \right), \qquad
\phi_c^2 \equiv \frac{2M^2\sqrt{\lambda}}{\gamma} .
\end{equation}
At $\phi^2>\phi_c^2$ the false vacuum is stable, while at $\phi^2<\phi_c^2$ the
effective mass-squared of $\chi$ becomes negative, and there is a tachyonic instability
leading to a rapid growth of $\chi$-modes and eventually to an end of the inflationary
expansion.

The evolution equations for the fields are given by
\begin{align}
\label{phi-eq}
\ddot \phi + 3 H \dot \phi - \nabla^2 \phi & = - \phi (m^2 + \gamma \chi^2),
\\
\label{chi-eq-new}
\ddot \chi+ 3 H \dot \chi - \nabla^2 \chi & =
(2\sqrt{2}M^2 - \gamma \phi^2 - \lambda \chi^2) \chi.
\end{align}
Before the waterfall transition, i.e., at $\phi^2>\phi_c^2$, the waterfall field is
trapped at $\chi=0$, so we can consider eq. (\ref{chi-eq-new}) as an equation
for the vacuum fluctuation $\delta \chi$.

One should note that, as we will see below, the typical scale of the fluctuations
amplified during the waterfall transition is of the order of the Hubble scale, or larger.
In such a case, the back-reaction effects (due to perturbations of metric) may
be non-negligible.
The taking into account the back-reaction effects, i.e., an using of the perturbed
FRW spacetime for the equation of motion for the scalar field $\phi$ at the stage
when the second field, $\chi$, is absent, can be most simply done, as is well
known (see, e.g., \cite{Bassett:2005xm, LL2009}), in the spatially flat gauge.
In this case, the perturbed equation coincides with the unperturbed one in the limit $\dot\phi=0$.
The smallness of $\dot\phi$ is provided by the slow-roll regime. So,
to minimize the back-reaction effects, we will work in the
vacuum-dominated regime, i.e., we assume that the vacuum energy $M^4$ dominates,
so that the Hubble parameter is effectively a constant,
\begin{equation}
H^2 = H_c^2 = \frac{M^4}{3 M_P^2},
\end{equation}
and one should consider only the case of small $\dot \phi$, satisfying the condition
$\dot\phi^2 \ll M_P^2 H_c^2$.

Another way to minimize back-reaction effects (and we will use it below) is
to choose the gauge with uniform $\phi$-slicing, in which $\delta\phi=0$.

As for eq. (\ref{chi-eq-new}), the back-reaction effects are small because the unperturbed
value of $\langle \chi \rangle$ is equal to zero \cite{Lyth:2010ch}.

The solution of eq. (\ref{phi-eq}) (in which we ignore
gradient term in accordance with our choice for the gauge) is
(for $t>t_c$, $t_c$ is the critical point when
the tachyonic instability begins)
\begin{equation}
\phi = \phi_c e^{-r H_c (t-t_c)}, \qquad r=\frac{3}{2} - \sqrt {\frac{9}{4}- \frac{m^2}{H_c^2}}.
\end{equation}

The scale factor $a$ is normalized to $1$ at
the time of the beginning of the waterfall (at $t=t_c=0$), so
\begin{equation}
a = e^{H_c t}.
\end{equation}
The conformal time in this case is
\begin{equation}
\eta = - \frac{1}{H_c} e^{-H_c t} = - \frac{1}{a H_c},
\end{equation}
corresponding to the de Sitter expansion.

\begin{figure}
\center
\includegraphics[trim = 0 0 0 0, width=8.5 cm]{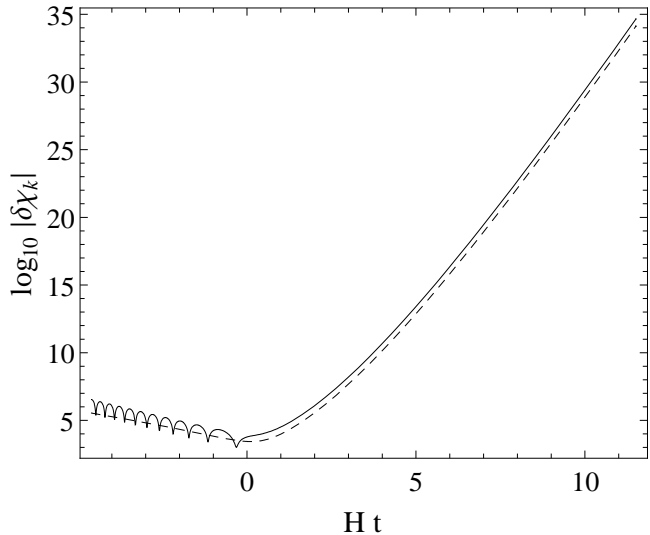} %
\caption{
The numerically calculated dependence of the module of $\delta\chi_k$ on time
for two values of $k$: $k=0.01 H_c$ (solid curve) and
$k=H_c$ (dashed curve). Other parameters used for the calculation are: $\beta=100$, $r=0.1$,
$H_c=10^{11}\;$GeV, $\phi_c=0.1 M_P$. For this set of parameters, the waterfall ends near
$H t \approx 3.5$, but we proceed the curves further just to show the asymptotic behavior.}
\label{fig-chit}
\end{figure}

From eq. (\ref{chi-eq-new}), one obtains the equation for Fourier modes of $\delta\chi$:
\begin{equation}
\label{deltaddotchik}
\delta \ddot\chi_k + 3 H \delta \dot\chi_k + \left(\frac{k^2}{a^2}-\beta H_c^2 +
\gamma \phi^2 \right) \delta \chi_k = 0 .
\end{equation}
Here, the parameter $\beta$ is given by the relation
\begin{equation}
\beta = 2 \sqrt{\lambda} \frac{M^2}{H_c^2}.
\end{equation}

Substituting the solution for $\phi$ in eq. (\ref{deltaddotchik}) and introducing the new
variable, $u=a\delta\chi$, one obtains the equation
\begin{equation}
u_k'' + (k^2 + \mu^2(\eta)) u_k = 0, \qquad
\mu^2(\eta) = \frac{\beta ( |\eta H_c|^{2 r} -1) -2 }{\eta^2} .
\end{equation}
Here, primes denote the derivative with respect to conformal time $\eta$.
The normalization at early times, when $k\gg\mu$, is $u = \frac{1}{\sqrt{2k}}e^{-ik\eta}$.
The definition of the power spectrum of $u$, as usual, is
\begin{equation}
{\cal P}_u = \frac{k^3}{2\pi^2} P_u = \frac{k^3}{2\pi^2} |u_k|^2.
\end{equation}

\begin{figure}
\center
\includegraphics[trim = 0 0 0 0, width=8.0 cm]{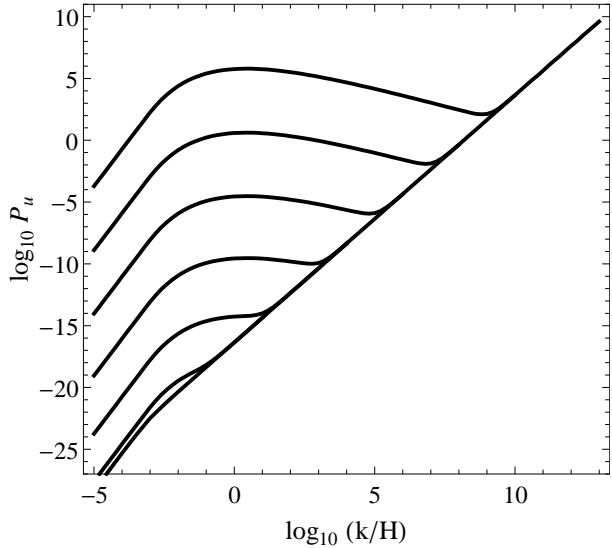} %
\caption{
The numerically calculated spectra ${\cal P}_{u}(k)$ at different moments of time,
for $\beta=1$, $r=0.1$, $H_c=10^{11}\;$GeV, $\phi_c=3\times 10^{-6} M_P$.
From bottom to top, $-\eta H = 10^3, 10^1, 10^{-1}, 10^{-3}, 10^{-5}, 10^{-7}, 10^{-9}$.
For these values of parameters, the waterfall ends at $-\eta H \approx 10^{-9}$.
}
\label{fig-Put}
\end{figure}

\begin{figure}
\center
\includegraphics[trim = 0 0 0 0, width=8.5 cm]{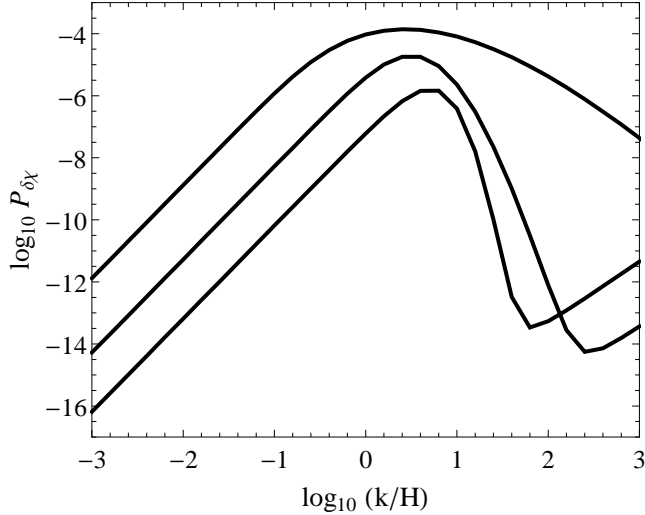} %
\caption{
The numerically calculated spectra ${\cal P}_{\delta \chi}(k)$ at the moment of the end of the waterfall.
For all cases, $r=0.1$, $H_c=10^{11}\;$GeV, $\phi_c=0.1 M_P$.
From bottom to top, $\beta=1500, 100, 7$.
}
\label{fig-Pchi}
\end{figure}

We show some results of numerical calculations in figures \ref{fig-chit}-\ref{fig-Pchi}.
In figure \ref{fig-chit} we
show how $\delta\chi_k$ depends on time for two different values of $k$ ($k=H$ and $k \ll H$).
It is seen that time evolution of both modes after the beginning of the waterfall is almost the same.
Following \cite{Lyth:2010zq} we assume that the growth era ends when the last term
in right-hand side of eq. (\ref{phi-eq}) becomes equal to the preceding one, i.e, when
\begin{equation}
\label{chinl}
\langle (\delta\chi)^2 \rangle  = \frac{m^2}{\gamma} \equiv \chi^2_{nl} .
\end{equation}

The asymptotic growth law of $k$-modes of the waterfall field is approximately \cite{Fonseca:2010nk}
\begin{equation}
\label{chi-beh}
\delta\chi_k \sim e^{s H t}, \qquad s = \sqrt{\frac{9}{4} + \beta} - \frac{3}{2}.
\end{equation}
It follows from eq. (\ref{chi-beh}) that at large $\beta$ the asymptotics of $\chi$
is $e^{\sqrt{\beta} H t}$, and the time scale of a period of the tachyonic instability
is much less than a Hubble time.

The typical example of the calculation of the power spectrum of $u$ is
shown in figure \ref{fig-Put}. This Figure is similar with figure 3 of \cite{Fonseca:2010nk},
where analogous quantities for the case
$\beta=100, r=0.1$ were shown. It is seen from our figure \ref{fig-Put}
that even for the case $\beta=1$ the waterfall is still effective, however it is much
slower (it takes $\sim 20$ e-folds for $\beta=1$ while in the case of $\beta=100$ the number
of e-folds is $\sim 3.5$).

Figure \ref{fig-Pchi} shows the form of ${\cal P}_{\delta \chi}(k)$ at the moment of the end of the
waterfall (determined by eq. (\ref{chinl})), for several parameter sets. Note also, that we
impose an artificial cutoff of large-$k$ modes in our numerical calculation, which corresponds
to considering only the waterfall field modes that already became classical at the beginning of
the waterfall. Technically, the cutoff is imposed at the
local minimum of ${\cal P}_{\delta \chi}(k)$-curve.

After a calculation of the power spectrum of $\delta \chi$, we calculate the spectrum ${P}_{(\delta\chi)^2}(k)$
at the end of the waterfall by the formula followed from the convolution theorem \cite{Lyth:1991ub}:
\begin{equation}
\label{Psvertka}
{ P}_{(\delta\chi)^2} (k) = \frac{2}{(2\pi)^3} \int d^3 k' P_{\delta\chi}(k') P_{\delta\chi}(| \vec k - \vec k'|).
\end{equation}

\section{ Curvature perturbation spectrum }
\label{sec-zeta}

The main equation for a calculation of the primordial curvature
perturbation (on uniform density hypersurfaces) is \cite{Wands:2000dp}
(see also \cite{GarciaBellido:1995qq, Lyth:1998xn})
\begin{equation}
\label{zetapnad}
\zeta = - \int dt \frac{H \delta p_{nad}}{p+\rho},
\end{equation}
where the non-adiabatic pressure perturbation is $\delta p_{nad}=\delta p - c_s^2 \delta\rho$
and the adiabatic sound speed is $c_s^2=\dot p / \dot \rho$. The formula (\ref{zetapnad})
follows from the ``separated universes'' picture
\cite{Starobinsky:1982ee, Starobinsky:1986fxa, Sasaki:1995aw, Wands:2000dp, Lyth:2004gb}
where, after smoothing over sufficiently
large scales, the universe becomes similar to an unperturbed FRW cosmology. In our
case, one has
\begin{equation}
\label{dpnad}
\delta p_{nad} = \delta p_{\chi} - \frac{\dot p}{\dot \rho} \; \delta \rho_{\chi}.
\end{equation}
Energy density $\rho$ and pressure $p$ is a sum of contributions of $\phi$ and $\chi$
fields. Eq. (\ref{dpnad}) takes into account that in $\delta p_{nad}$ there is
no contribution from $\phi$ field.

Everywhere below we will use for the fluctuations of the $\chi$ field the more simple
notation, than that is used in the previous section: $\chi$ instead
of $\delta\chi$. It is just the same because $\langle \chi \rangle =0$.

Our numerical calculations of $\chi_k(t)$ in section \ref{sec-wf} (two examples are presented
in figure \ref{fig-chit}) show that a form of the $t$-dependence of $\chi_k$ weakly depends on
$k$. Therefore, we assume that, approximately, the time dependence of $\chi({\bf x}, t)$ is
the same as of $\chi_k(t)$ at $k\sim k_*$, where $k_*$ is the peak value of ${\cal P}_\chi$.
We designate this dependence (with arbitrary normalization) by $f(t)$ (this function
depends on the model parameters $\beta, r$). Then, one has
\begin{equation}
\label{chixt}
\chi({\bf x}, t) = C({\bf x}) f(t),
\end{equation}
i.e., the dependence on a location is factorized, and
\begin{equation}
\dot \chi({\bf x}, t) = C({\bf x}) \dot f(t) = \chi({\bf x}, t) \frac{\dot f(t)}{f(t)}.
\end{equation}

Expressions for the energy density and pressure of the $\chi$ field are
\begin{align}
\label{pchiwithgrad}
p_\chi ({\bf x}, t)& = - \frac{m_\chi^2(t)}{2} \chi^2({\bf x}, t) + \frac{1}{2} \dot \chi^2 ({\bf x}, t)-
\frac{1}{6 a^2} \left| \nabla \chi ({\bf x}, t)\right|^2,
\\
\label{rhochiwithgrad}
\rho_\chi ({\bf x}, t)& = \frac{m_\chi^2(t)}{2} \chi^2({\bf x}, t) + \frac{1}{2} \dot \chi^2({\bf x}, t) +
\frac{1}{2 a^2} \left| \nabla \chi ({\bf x}, t) \right|^2
\end{align}
(where, for brevity, $m_\chi^2\left(\phi(t)\right) \equiv m_\chi^2(t)$).

For calculations of the curvature perturbation spectra we will use the spatially averaged
energy density and pressure, i.e., we do in equations (\ref{pchiwithgrad}, \ref{rhochiwithgrad})
the substitutions $\chi^2 \to \langle \chi^2 \rangle$,
$\dot \chi^2 \to \langle \dot \chi^2 \rangle$,
$|\nabla \chi|^2 \to \langle | \nabla \chi | ^2 \rangle$.

We proved the relative smallness of gradient terms in these expressions using the approximate
method suggested in \cite{Lyth:2010zq} for the case when cosmological expansion during the waterfall
is ignored. Namely, due to a peak feature in ${\cal P}_\chi(k)$-dependence (see figure \ref{fig-Pchi})
one has, approximately,
\begin{equation}
\label{intgrad}
\langle \left| \nabla \chi ({\bf x}, t) \right|^2 \rangle \sim \int d^3 k k^2 |\chi_k|^2
\sim k_*^2 (t) \langle \chi^2 (t) \rangle.
\end{equation}

For values of parameters we operate with, $k_* \sim H_c$. At the end of the waterfall ($t=t_{end}$),
the estimate for the gradient term will be
\begin{equation}
\langle \left| \nabla \chi \right|^2 \rangle \sim H_c^2 \chi_{nl}^2,
\end{equation}
and the relation between gradient and time derivative terms in (\ref{rhochiwithgrad})
(we denote it with letter $D$) can be written as
\begin{equation}
D \equiv \frac {\langle \left| \nabla \chi \right|^2 \rangle } {a^2 \langle \dot \chi^2 \rangle}
\sim \frac{1}{s^2 a(t_{end})^2},
\end{equation}
where we have used the asymptotic behavior (\ref{chi-beh}) for this estimate.
For $\beta \gg 1$, $D$ is negligible due to the large factor $s$, $s \gg 1$. For $\beta \sim 1$,
$s\sim 1$, but $a \gg 1$. We conclude that $D$ is always small and the contribution
of spatial gradient in (\ref{pchiwithgrad}, \ref{rhochiwithgrad}) can be always
neglected (note that in case when $\beta$ is small, the peak in ${\cal P}_{\delta\chi}(k)$,
as seen from figures \ref{fig-Put} and \ref{fig-Pchi}, is very broad, so in this case the smallness of gradient
terms is proved by the straightforward calculation of (\ref{intgrad})).

For a calculation of $\delta p_{nad}$ we need expressions for fluctuations of energy density
and pressure. They are given by
\begin{align}
\delta p_\chi & = \frac{1}{2} \left[- m_\chi^2(t) + \left( \frac{\dot f(t)}{f(t)} \right)^2 \right]\delta \chi^2,
\\
\delta \rho_\chi & = \frac{1}{2} \left[ m_\chi^2(t) +  \left( \frac{\dot f(t)}{f(t)} \right)^2 \right]\delta \chi^2,
\end{align}
where $\delta \chi^2 = \chi^2 - \langle \chi^2 \rangle$.
Averaged values of $\chi^2$ and $\dot \chi^2$ are calculated taking into
account the equation (\ref{chinl}):
\begin{align}
\langle \chi^2(t) \rangle & = \chi_{nl}^2 \frac{f^2(t)}{f^2(t_{end})},
\\
\langle \dot \chi^2(t) \rangle & = \chi_{nl}^2 \frac{\dot f^2(t)}{f^2(t_{end})}.
\end{align}
Here, $t_{end}$ is a moment of time of an end of the waterfall.
Finally, for the average values of the energy density and pressure, one has
\begin{align}
\langle p_\chi \rangle & = \frac{1}{2} \left[- m_\chi^2(t) +  \left( \frac{\dot f(t)}{f(t)} \right)^2 \right]
\left( \frac{f(t)}{f(t_{end})} \right)^2 \chi_{nl}^2,
\\
\langle \rho_\chi \rangle & = \frac{1}{2} \left[ m_\chi^2(t) +  \left( \frac{\dot f(t)}{f(t)} \right)^2 \right]
\left( \frac{f(t)}{f(t_{end})} \right)^2 \chi_{nl}^2.
\end{align}

The values of $p_\phi$ and $\rho_\phi$ are
\begin{equation}
p_\phi = -V(\phi) + \frac{1}{2} \dot\phi^2, \qquad
 \rho_\phi = V(\phi) + \frac{1}{2} \dot\phi^2,
\end{equation}
where, for our case, $V(\phi) = \frac{1}{2} m^2 \phi^2$.

For the curvature perturbation, we have the integral
\begin{equation} \label{zeta-minus}
\zeta=\zeta_\chi=-\int \frac{H_c dt}{\dot\phi^2(t) + \langle \dot\chi^2(t) \rangle}
 \frac{\delta \chi^2(t)}{2} \left[- m_\chi^2(t) + \left( \frac{\dot f(t)}{f(t)} \right)^2  -
  \frac{\dot p}{\dot \rho}
  \left( m_\chi^2(t) + \left( \frac{\dot f(t)}{f(t)} \right)^2 \right) \right].
\end{equation}
The relation between curvature perturbation and $\chi^2$-spectra can be written as
\begin{equation}
\label{PA}
{\cal P}_\zeta = A^2 {\cal P}_{\delta \chi^2} (t_{end}) ,
\end{equation}
where the value of the spectrum in the right-hand side is calculated at a time of the end of the waterfall,
$t_{end}$, and $A$ is to be determined.

It is easy to check that for a calculation of the spectrum ${\cal P}_{\delta \chi^2}$ in
equation (\ref{PA}) one can use the expression for  ${\cal P}_{\chi^2}$ given by the
convolution formula (\ref{Psvertka}).

The time dependence of $\chi^2(t)$ needed for the calculation of the integral in
(\ref{zeta-minus}) is given by
\begin{equation}
\chi^2(t) = \chi^2(t_{end}) \left( \frac{f(t)}{f(t_{end})} \right)^2.
\end{equation}
Using this relation, we can extract the value of $A$:
\begin{equation} A = \int\limits_0^{t_{end}} \frac{H_c dt}{\dot \phi^2(t) + \langle\dot \chi^2(t)\rangle}
\left( \frac{f(t)}{f(t_{end}) } \right)^2
 \frac{1}{2} \left[- m_\chi^2(t) +  \left( \frac{\dot f(t)}{f(t)} \right)^2  -
  \frac{\dot p}{\dot \rho}
  \left( m_\chi^2(t) +  \left( \frac{\dot f(t)}{f(t)} \right)^2 \right) \right].
\end{equation}

\begin{figure}
\center
\includegraphics[trim = 0 0 0 0, width=8.8 cm]{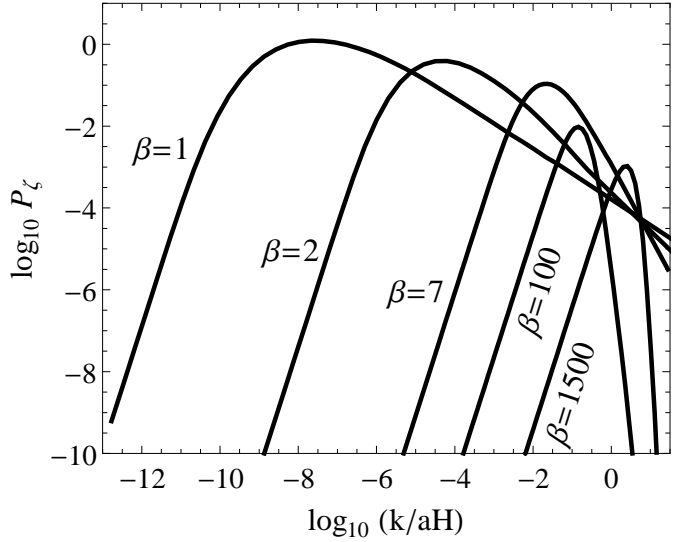} %
\caption{
The numerically calculated spectra ${\cal P}_{\zeta}(k)$ at the moment of the end of the waterfall.
Parameters used for the calculation are:  $r=0.1$, $H_c=10^{11}\;$GeV (all curves);
$\phi_c=3\times 10^{-6} M_P$ (for $\beta=1, 2, 7)$; $\phi_c=0.1 M_P$ (for $\beta=100, 1500)$.
}
\label{fig-Pzeta}
\end{figure}

We show results of the calculation of ${\cal P}_\zeta(k)$ in figure \ref{fig-Pzeta}.
It is seen that the spectrum can reach rather large values (of order of 1) for
the case of $\beta\sim 1$ (if $r$ is in a broad interval, say, $0.1 \div 0.001$).

\begin{figure}
\includegraphics[trim = 0 0 0 0, width=7.5 cm]{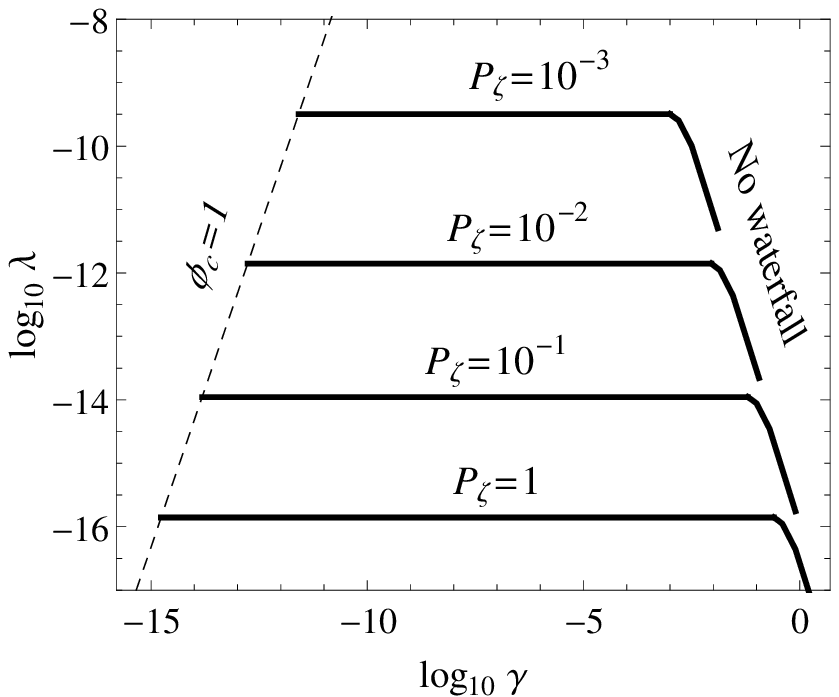} %
$\;\;\;$
\includegraphics[trim = 0 0 0 0, width=7.5 cm]{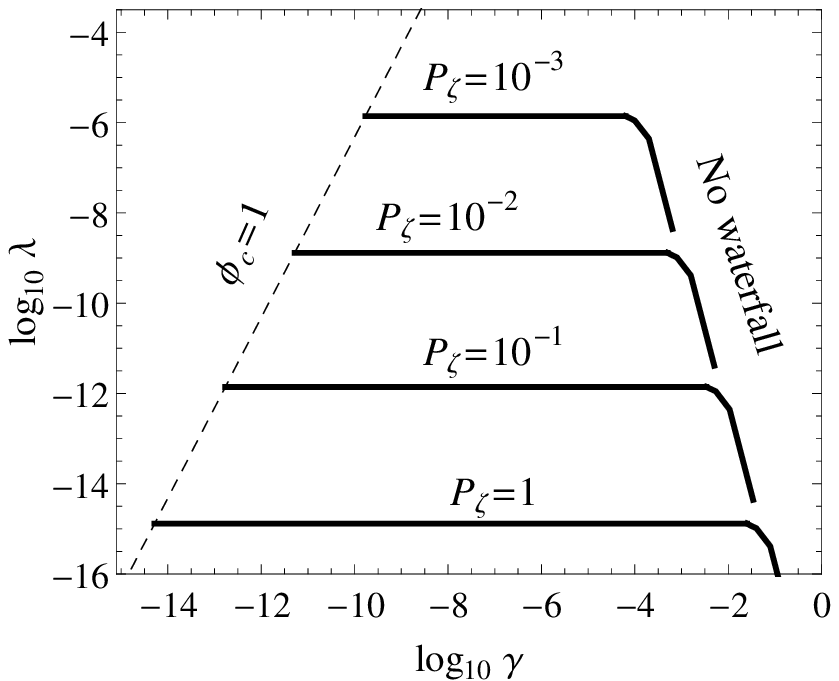} %
\caption{
The allowed ranges for the values of $\gamma$ and $\lambda$ and the corresponding
maximum values of ${\cal P}_\zeta(k)$ that are reached. Left panel: $r=0.1$ ($m\approx 0.5 H_c$).
Right panel: $r=0.001$ ($m\approx 0.05 H_c$). For both panels, $H_c=10^{11}\;$GeV.
}
\label{fig-constr}
\end{figure}

In figure \ref{fig-constr} we show the model parameter regions that lead to rather large
values of ${\cal P}_\zeta$. We also show, on the same Figure, the parameter region where
waterfall is not effective (classical regime is not reached) and boundary of
$\phi_c=1$ (in units of Planck mass $M_P$).

At the end of this section, we numerate the constraints on the parameter space which
follow from the assumptions used for an obtaining of the main results.

\begin{enumerate}
\item
Slow roll of the inflaton field:
\begin{equation}
m^2 \ll H_c^2.
\end{equation}

\item
False vacuum dominance:
\begin{equation}
M^4 \gg \frac{1}{2} m^2 \phi_c^2.
\end{equation}

\item
Small $\phi$-condition (for justifying of the omission of high powers of $\phi$
in the inflaton potential):
\begin{equation}
\phi_c \ll M_P.
\end{equation}

\item
Small effects from back-reaction (slow-roll condition):
\begin{equation}
\dot\phi^2 \ll M_P^2 H^2.
\end{equation}

\item
Condition for $\delta \chi_k$ to be classical field \cite{Lyth:2010ch}
\begin{equation}
\sqrt{\gamma} \ll \frac{1}{\sqrt{\beta}}.
\end{equation}

\item
Condition for a dominance of the scenario with the inflaton's classical
rolling \cite{Dufaux:2008dn}:
\begin{equation}
\dot \phi_c > \gamma \lambda v^2, \quad {\rm or} \quad \gamma^{3/2} < \sqrt{r}.
\end{equation}

\end{enumerate}

\section{PBH production from non-Gaussian perturbations}
\label{sec-pbh}

A production of PBHs (about these objects, see, e.g., reviews \cite{Khlopov:2008qy, Carr:2009jm})
during reheating process had been considered in works
\cite{Bassett:2000ha, Green:2000he, Suyama:2004mz, Suyama:2006sr, Finelli:2000gi}.
The classical PBH formation criterion in the radiation-dominated epoch is \cite{Carr:1974nx}
\begin{equation}
\label{limit-delta}
\delta > \delta_c \approx 1/3,
\end{equation}
where $\delta$ is the smoothed density contrast at horizon crossing.
The Fourier component of the comoving density perturbation $\delta$ is related to the Fourier
component of the Bardeen potential $\Psi$ as
\begin{equation}
\delta_k = - \frac{2}{3} \left( \frac{k}{aH} \right)^2 \Psi_k.
\end{equation}
For modes in a super-horizon regime, $\Psi_k \approx -(2/3) {\cal R}_k \approx -(2/3) \zeta_k$,
so (\ref{limit-delta}) can be translated to a limiting value of the curvature perturbation
\cite{Hidalgo:2007vk}, which is
\begin{equation}
\label{zetath}
\zeta_{c} \approx 0.7.
\end{equation}

It is seen from (\ref{zeta-minus}) that the curvature perturbation generated by waterwall field
has a negative sign, so, naively, the threshold (\ref{zetath}) can't be reached.
However, the perturbations must be considered with respect to the average value, so
\begin{equation}
\zeta \to \zeta_0 = \zeta - \langle \zeta \rangle,
\end{equation}
and the condition for individual PBH formation is $\zeta_0 \gtrsim \zeta_c$,
while PBHs will exceed the currently available limits on their average abundance already for
$| \langle \zeta \rangle |$ only slightly exceeding $\zeta_c$ (i.e.,
$| \langle \zeta \rangle |-\zeta_c \ll 1$) \cite{Lyth:2011kj}, so, practically,
the PBH constraint on $\zeta$ is just $ | \langle \zeta \rangle | < \zeta_c$.

For the distribution of $\zeta_0$ we may write
\begin{equation}
\zeta_0= - (g^2 - \langle g^2 \rangle ),
\end{equation}
where $g$ is gaussian (in our case, $g \sim \delta \chi$), and for the average of $\zeta_0^2$,
using known properties of Gaussian distributions,
\begin{equation}
\langle \zeta_0^2 \rangle = 2 \langle g^2 \rangle^2 = 2 | \langle \zeta \rangle |^2,
\end{equation}
so, in terms of $\langle \zeta_0^2 \rangle$ the significant PBH formation will happen if
\begin{equation}
\label{lim-zeta2}
\langle \zeta_0^2 \rangle = \int {\cal P}_\zeta(k) \frac{dk}{k}  \; \gtrsim \; 2 \zeta_{c}^2 \approx 1.
\end{equation}

In case of ${\cal P}_\zeta$-spectra shown in figure \ref{fig-Pzeta}, the curve corresponding to
$\beta=2$ is close to the bound (\ref{lim-zeta2}) while the curve for $\beta=1$ has
$\langle \zeta_0^2 \rangle \approx 7$ which makes that set of parameters forbidden by the
PBH formation constraint.

The mass of the PBHs produced from curvature perturbation spectra presented in figure \ref{fig-Pzeta}
can be estimated as follows. The horizon mass at the end of inflation is
\begin{equation}
\label{Mi}
M_i \approx (H_c^{-1})^3 \rho = \frac{3 M_P^2}{H_c}  \sim 10^2 \; {\rm g},
\end{equation}
and using a well-known dependence for the horizon mass corresponding to
fluctuation having wave number $k$, $M_h \sim k^{-2}$ (see, e.g., \cite{Bugaev:2008gw}),
we have
\begin{equation}
\label{Mh}
M_{BH} \approx M_h \approx M_i \left( \frac{a(t_{end}) H_c}{k} \right)^2.
\end{equation}
For the case of figure \ref{fig-Pzeta}, curve for $\beta=1$ corresponds to
$M_{BH}\sim 10^{19}\;$g (the mass range of non-evaporating PBHs that can constitute dark
matter) while $\beta=2$ corresponds to $M_{BH}\sim 10^{13}\;$g (such PBHs have already
evaporated, but the products of their evaporation are, in principle, observable).

\begin{figure}
\center
\includegraphics[trim = 0 0 0 0, width=8.8 cm]{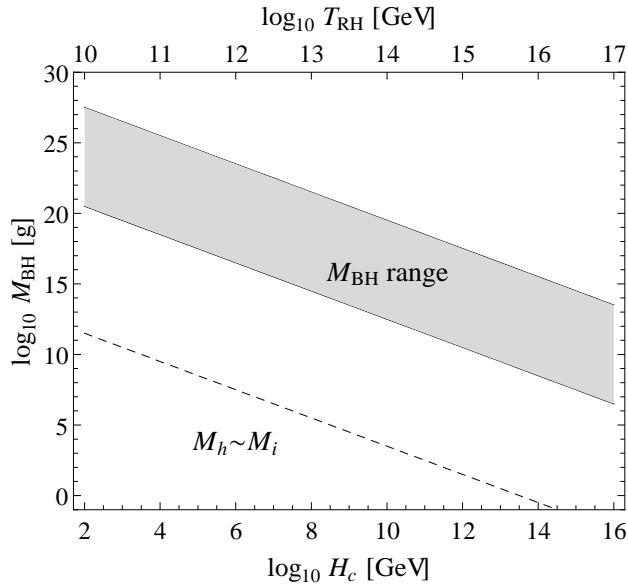} %
\caption{
The range of characteristic masses of PBHs that can be produced by a hybrid inflation
waterfall, as a function of inflation energy scale or reheating temperature.
}
\label{fig-BHmasses}
\end{figure}

For other values of inflation energy scale, or reheating temperature, which
is connected to $H_c$ by the relation \cite{Bugaev:2008gw}
\begin{equation}
\label{TRH}
T_{RH} = \left( \frac{90 M_P^2 H_c^2} {\pi^2 g_*} \right)^{1/4}, \qquad g_* \approx 100,
\end{equation}
we sketch the expected range of characteristic masses of PBHs, that can be produced
(see figure \ref{fig-BHmasses}).
The shaded region in the figure corresponds to large values of $\beta$ parameter
($ 1 \lesssim \beta \lesssim  2$).
The results shown in this figure must be considered together
with the particular cosmological constraints on PBH abundance.
We leave this question for further studies.

\section {Conclusions}
\label{sec-concl}

We carried out numerical calculations of a contribution of the waterfall field to the
primordial curvature perturbation (on uniform density hypersurfaces) $\zeta$, which is produced
during waterfall transition in hybrid inflation scenario. The calculation is performed
for a broad interval of values of the model parameters.

We did not consider the contribution to $\zeta$ from inflaton rolling, which is
dominant at cosmological scales. One should note that the simple quadratic inflationary
potential used in the present paper can be easily corrected (see, e.g., \cite{Rehman:2009wv})
to give a red-tilted spectrum at cosmological scales (converting, e.g., the original hybrid
inflation model to a hilltop model \cite{Boubekeur:2005zm, Kohri:2007gq}), without any
modification of our predictions for small scales.

Main results of the paper are shown in figures \ref{fig-Pzeta},
\ref{fig-constr}, \ref{fig-BHmasses}. One can
see from the figure \ref{fig-Pzeta} that peak amplitudes of ${\cal P}_\zeta$ strongly
depend on the value of $\beta$ (curiously, ${\cal P}_\zeta \sim 1$ corresponds to
$\beta\sim 1$ in a broad interval of $\gamma$ and $r$). The peak values, $k_*$, for
small $\beta$ are far beyond horizon, so, the smoothing over the horizon size will
not decrease the peak values of the smoothed spectrum. Furthermore, the spectrum
near peak remains strongly non-Gaussian after the smoothing. Our calculations based
on the quadratic inflaton potential show that for $\beta \lesssim 100$ and in the broad
interval of $r$ the peak value $k_*$ can be estimated by the simple relation:
\begin{equation}
\frac{k_*}{aH} \sim e^{-N},
\end{equation}
where $N$ is the number of e-folds during the waterfall transition. The similar estimate
is contained in the recent work \cite{Lyth:2011kj}.

We conclude that the strong growth of amplitudes of the curvature perturbation spectrum
at $\beta \to 1$, which had been anticipated in pioneering works
\cite{Randall:1995dj, GarciaBellido:1996qt}, really takes place. However,
the condition which is very often used as a constraint, that the bare mass-squared of
the $\chi$-field, $-m_\chi^2 = 2\sqrt{\lambda}M^2$, must be \emph{much larger}
than $H^2$ (the so-called ``waterfall condition'' \cite{Copeland:1994vg}) seems
to be too restrictive. Our results, presented in figures \ref{fig-Pzeta} and \ref{fig-constr},
show that only at $-m_\chi^2 \approx H^2$ (when $\beta \sim 1$) the
amplitude of the curvature spectrum ${\cal P}_\zeta$ becomes close to one, i.e.,
enters a region that can be constrained by PBH data. In figure
\ref{fig-BHmasses} we show the region of PBH masses
which can be produced due to the waterfall transition in hybrid inflation,
in the case of large $\beta$ parameter. Abundances of these PBHs should be constrained
by data of PBH searches.
It would be very interesting to analyze the corresponding constraints following from data
of relict GW searches (see, e.g., \cite{Dufaux:2008dn, Bugaev:2010bb, Giovannini:2010tk}).

% \paragraph*{Acknowledgments.}
% The work was ...

\acknowledgments

The authors are grateful to Profs. D.~H.~Lyth and A.~A.~Starobinsky for important remarks.

 %{references}


\begin{thebibliography}{99} %{references}

\bibitem{Lyth:2010ch}
  D.~H.~Lyth,
  {\it ``Issues concerning the waterfall of hybrid inflation,''}
  arXiv:1005.2461.
  %%CITATION = ARXIV:1005.2461;%%

\bibitem{Gong:2010zf}
  J.~O.~Gong and M.~Sasaki,
  {\it ``Waterfall field in hybrid inflation and curvature perturbation,''}
  JCAP {\bf 1103} (2011) 028  [arXiv:1010.3405].
  %%CITATION = JCAPA,1103,028;%%

\bibitem{Fonseca:2010nk}
  J.~Fonseca, M.~Sasaki and D.~Wands,
  {\it ``Large-scale Perturbations from the Waterfall Field in Hybrid Inflation,''}
  JCAP {\bf 1009} (2010) 012  [arXiv:1005.4053].
  %%CITATION = JCAPA,1009,012;%%

\bibitem{Abolhasani:2010kr}
  A.~A.~Abolhasani and H.~Firouzjahi,
  {\it ``No Large Scale Curvature Perturbations during Waterfall of Hybrid Inflation,''}
  Phys.\ Rev.\  {\bf D 83} (2011) 063513 [arXiv:1005.2934].

\bibitem{Abolhasani:2010kn}
  A.~A.~Abolhasani, H.~Firouzjahi and M.~H.~Namjoo,
  {\it ``Curvature Perturbations and non-Gaussianities from Waterfall Phase Transition during Inflation,''}
  Class.\ Quant.\ Grav.\  {\bf 28} (2011) 075009 [arXiv:1010.6292].

\bibitem{Lyth:2010zq}
  D.~H.~Lyth,
  {\it ``Contribution of the hybrid inflation waterfall to the primordial curvature perturbation,''}
  JCAP {\bf 1107} (2011) 035 [arXiv:1012.4617].

\bibitem{Abolhasani:2011yp}
  A.~A.~Abolhasani, H.~Firouzjahi and M.~Sasaki,
  {\it ``Curvature perturbation and waterfall dynamics in hybrid inflation,''}
  arXiv:1106.6315.

\bibitem{Lyth:2011kj}
  D.~H.~Lyth,
  {\it ``Primordial black hole formation and hybrid inflation,''}
  arXiv:1107.1681.

\bibitem{Barnaby:2006cq}
  N.~Barnaby and J.~M.~Cline,
  {\it ``Nongaussian and nonscale-invariant perturbations from tachyonic preheating in hybrid inflation,''}
  Phys.\ Rev.\  {\bf D 73} (2006) 106012 [astro-ph/0601481].

\bibitem{Barnaby:2006km}
  N.~Barnaby and J.~M.~Cline,
  {\it ``Nongaussianity from Tachyonic Preheating in Hybrid Inflation,''}
  Phys.\ Rev.\  {\bf D 75} (2007) 086004 [astro-ph/0611750].

\bibitem{Kallosh:2007ig}
  R.~Kallosh,
  {\it ``On inflation in string theory,''}
  Lect.\ Notes Phys.\  {\bf 738} (2008) 119-156 [hep-th/0702059].

\bibitem{Kallosh:2003ux}
  R.~Kallosh and A.~D.~Linde,
  {\it ``P term, D term and F term inflation,''}
  JCAP {\bf 0310} (2003) 008 [hep-th/0306058].

\bibitem{Traschen:1990sw}
  J.~H.~Traschen and R.~H.~Brandenberger,
  {\it ``Particle Production During Out-of-equilibrium Phase Transitions,''}
  Phys.\ Rev.\  {\bf D 42} (1990) 2491-2504.

\bibitem{Dolgov:1989us}
  A.~D.~Dolgov and D.~P.~Kirilova,
  {\it ``On Particle Creation By A Time Dependent Scalar Field,''}
  Sov.\ J.\ Nucl.\ Phys.\  {\bf 51} (1990) 172-177 .

\bibitem{Shtanov:1994ce}
  Y.~Shtanov, J.~H.~Traschen and R.~H.~Brandenberger,
  {\it ``Universe reheating after inflation,''}
  Phys.\ Rev.\  {\bf D 51} (1995) 5438-5455 [hep-ph/9407247].

\bibitem{Kofman:1994rk}
  L.~Kofman, A.~D.~Linde and A.~A.~Starobinsky,
  {\it ``Reheating after inflation,''}
  Phys.\ Rev.\ Lett.\  {\bf 73} (1994) 3195 [hep-th/9405187].

\bibitem{Felder:2000hj}
  G.~N.~Felder, J.~Garcia-Bellido, P.~B.~Greene, L.~Kofman, A.~D.~Linde and I.~Tkachev,
  {\it ``Dynamics of symmetry breaking and tachyonic preheating,''}
  Phys.\ Rev.\ Lett.\  {\bf 87} (2001) 011601 [hep-ph/0012142].

\bibitem{Bassett:1998wg}
  B.~A.~Bassett, D.~I.~Kaiser, R.~Maartens,
  {\it ``General relativistic preheating after inflation,''}
  Phys.\ Lett.\  {\bf B 455} (1999) 84-89 [hep-ph/9808404].

\bibitem{Liddle:1999hq}
  A.~R.~Liddle, D.~H.~Lyth, K.~A.~Malik, D.~Wands,
  {\it ``Superhorizon perturbations and preheating,''}
  Phys.\ Rev.\  {\bf D 61} (2000) 103509 [hep-ph/9912473].

\bibitem{Green:2000he}
  A.~M.~Green and K.~A.~Malik,
  {\it ``Primordial black hole production due to preheating,''}
  Phys.\ Rev.\  {\bf D 64} (2001) 021301 [hep-ph/0008113].


\bibitem{Pilo:2004ke}
  L.~Pilo, A.~Riotto and A.~Zaffaroni,
  {\it ``On the amount of gravitational waves from inflation,''}
  Phys.\ Rev.\ Lett.\  {\bf 92} (2004) 201303 [astro-ph/0401302].
  %%CITATION = PRLTA,92,201303;%%

\bibitem{Gong:2008ni}
  J.~-O.~Gong and M.~Sasaki,
  {\it ``Curvature perturbation spectrum from false vacuum inflation,''}
  JCAP {\bf 0901} (2009) 001 [arXiv:0804.4488].

\bibitem{Suyama:2004mz}
  T.~Suyama, T.~Tanaka, B.~Bassett and H.~Kudoh,
  {\it ``Are black holes over-produced during preheating?,''}
  Phys.\ Rev.\  {\bf D 71} (2005) 063507 [hep-ph/0410247].
  %%CITATION = PHRVA,D71,063507;%%

\bibitem{Suyama:2006sr}
  T.~Suyama, T.~Tanaka, B.~Bassett and H.~Kudoh,
  {\it ``Black hole production in tachyonic preheating,''}
  JCAP {\bf 0604} (2006) 001 [hep-ph/0601108].
  %%CITATION = JCAPA,0604,001;%%

\bibitem{Brax:2010ai}
  P.~Brax, J.~-F.~Dufaux and S.~Mariadassou,
  {\it ``Preheating after Small-Field Inflation,''}
  Phys.\ Rev.\  {\bf D 83} (2011) 103510 [arXiv:1012.4656].

\bibitem{Dufaux:2008dn}
  J.~F.~Dufaux, G.~N.~Felder, L.~Kofman and O.~Navros,
  {\it ``Gravity Waves from Tachyonic Preheating after Hybrid Inflation,''}
  JCAP {\bf 0903} (2009) 001 [arXiv:0812.2917].

\bibitem{Kofman:1997yn}
  L.~Kofman, A.~D.~Linde and A.~A.~Starobinsky,
  {\it ``Towards the theory of reheating after inflation,''}
  Phys.\ Rev.\  {\bf D 56} (1997) 3258 [hep-ph/9704452].
  %%CITATION = PHRVA,D56,3258;%%

\bibitem{Asaka:2001ez}
  T.~Asaka, W.~Buchmuller and L.~Covi,
  {\it ``False vacuum decay after inflation,''}
  Phys.\ Lett.\  {\bf B 510} (2001) 271-276 [hep-ph/0104037].

\bibitem{Copeland:2002ku}
  E.~J.~Copeland, S.~Pascoli and A.~Rajantie,
  {\it ``Dynamics of tachyonic preheating after hybrid inflation,''}
  Phys.\ Rev.\  {\bf D 65} (2002) 103517 [hep-ph/0202031].
  %%CITATION = PHRVA,D65,103517;%%

\bibitem{GarciaBellido:2002aj}
  J.~Garcia-Bellido, M.~Garcia Perez and A.~Gonzalez-Arroyo,
  {\it ``Symmetry breaking and false vacuum decay after hybrid inflation,''}
  Phys.\ Rev.\  D {\bf 67} (2003) 103501 [hep-ph/0208228].
  %%CITATION = PHRVA,D67,103501;%%

\bibitem{Starobinsky:1982ee}
  A.~A.~Starobinsky,
  {\it ``Dynamics of Phase Transition in the New Inflationary Universe Scenario and Generation of Perturbations,''}
  Phys.\ Lett.\  {\bf B117} (1982) 175-178.

\bibitem{Starobinsky:1986fxa}
  A.~A.~Starobinsky,
  {\it ``Multicomponent de Sitter (Inflationary) Stages and the Generation of Perturbations,''}
  JETP Lett.\  {\bf 42} (1985) 152-155.

\bibitem{Sasaki:1995aw}
  M.~Sasaki and E.~D.~Stewart,
  {\it ``A General analytic formula for the spectral index of the density perturbations produced during inflation,''}
  Prog.\ Theor.\ Phys.\  {\bf 95} (1996) 71 [astro-ph/9507001].

\bibitem{Lyth:2004gb}
  D.~H.~Lyth, K.~A.~Malik and M.~Sasaki,
  {\it ``A General proof of the conservation of the curvature perturbation,''}
  JCAP {\bf 0505} (2005) 004 [astro-ph/0411220].

\bibitem{Linde:1991km}
  A.~D.~Linde,
  {\it ``Axions in inflationary cosmology,''}
  Phys.\ Lett.\  B {\bf 259} (1991) 38.

\bibitem{Linde:1993cn}
  A.~D.~Linde,
  {\it ``Hybrid inflation,''}
  Phys.\ Rev.\  {\bf D 49} (1994) 748-754 [astro-ph/9307002].

\bibitem{Bassett:2005xm}
  B.~A.~Bassett, S.~Tsujikawa, D.~Wands,
  {\it ``Inflation dynamics and reheating,''}
  Rev.\ Mod.\ Phys.\  {\bf 78} (2006) 537-589 [astro-ph/0507632].

\bibitem{LL2009} D.~H.~Lyth and A.~R.~Liddle,
 {\it ``The primordial density perturbation,''}
 Cambridge University Press (2009).

\bibitem{Lyth:1991ub}
  D.~H.~Lyth,
  {\it ``Axions and inflation: Sitting in the vacuum,''}
  Phys.\ Rev.\  {\bf D 45} (1992) 3394-3404.

\bibitem{Wands:2000dp}
  D.~Wands, K.~A.~Malik, D.~H.~Lyth and A.~R.~Liddle,
  {\it ``A New approach to the evolution of cosmological perturbations on large scales,''}
  Phys.\ Rev.\ {\bf D 62} (2000) 043527 [astro-ph/0003278].

\bibitem{Lyth:1998xn}
  D.~H.~Lyth, A.~Riotto,
  {\it ``Particle physics models of inflation and the cosmological density perturbation,''}
  Phys.\ Rept.\  {\bf 314} (1999) 1-146 [hep-ph/9807278].

\bibitem{GarciaBellido:1995qq}
  J.~Garcia-Bellido, D.~Wands,
  {\it ``Metric perturbations in two field inflation,''}
  Phys.\ Rev.\  {\bf D 53} (1996) 5437-5445 [astro-ph/9511029].


% PBH sec.

\bibitem{Khlopov:2008qy}
  M.~Y.~Khlopov,
  {\it ``Primordial Black Holes,''}
  Res.\ Astron.\ Astrophys.\  {\bf 10} (2010) 495-528 [arXiv:0801.0116].

\bibitem{Carr:2009jm}
  B.~J.~Carr, K.~Kohri, Y.~Sendouda and J.~Yokoyama,
  {\it ``New cosmological constraints on primordial black holes,''}
  Phys.\ Rev.\  {\bf D 81} (2010) 104019 [arXiv:0912.5297].

\bibitem{Bassett:2000ha}
  B.~A.~Bassett and S.~Tsujikawa,
  {\it ``Inflationary preheating and primordial black holes,''}
  Phys.\ Rev.\  {\bf D 63} (2001) 123503 [hep-ph/0008328].

\bibitem{Finelli:2000gi}
  F.~Finelli and S.~Khlebnikov,
  {\it ``Large metric perturbations from rescattering,''}
  Phys.\ Lett.\  {\bf B 504} (2001) 309-313 [hep-ph/0009093].

\bibitem{Carr:1974nx}
  B.~J.~Carr and S.~W.~Hawking,
  {\it ``Black holes in the early Universe,''}
  Mon.\ Not.\ Roy.\ Astron.\ Soc.\  {\bf 168} (1974) 399-415.

\bibitem{Hidalgo:2007vk}
  J.~C.~Hidalgo,
  {\it ``The effect of non-Gaussian curvature perturbations on the formation of primordial black holes,''}
  arXiv:0708.3875.

\bibitem{Bugaev:2008gw}
  E.~Bugaev and P.~Klimai,
  {\it ``Constraints on amplitudes of curvature perturbations from primordial black holes,''}
  Phys.\ Rev.\  {\bf D 79} (2009) 103511 [arXiv:0812.4247].
  %%CITATION = PHRVA,D79,103511;%%

% Conclusions

\bibitem{Rehman:2009wv}
  M.~U.~Rehman, Q.~Shafi and J.~R.~Wickman,
  {\it ``Hybrid Inflation Revisited in Light of WMAP5,''}
  Phys.\ Rev.\  {\bf D 79} (2009) 103503 [arXiv:0901.4345].

\bibitem{Boubekeur:2005zm}
  L.~Boubekeur and D.~H.~Lyth,
  {\it ``Hilltop inflation,''}
  JCAP {\bf 0507} (2005) 010 [hep-ph/0502047].

\bibitem{Kohri:2007gq}
  K.~Kohri, C.-M.~Lin and D.~H.~Lyth,
  {\it ``More hilltop inflation models,''}
  JCAP {\bf 0712} (2007) 004 [arXiv:0707.3826].

\bibitem{Randall:1995dj}
  L.~Randall, M.~Soljacic and A.~H.~Guth,
  {\it ``Supernatural inflation: Inflation from supersymmetry with no (very) small parameters,''}
  Nucl.\ Phys.\  {\bf B 472} (1996) 377-408 [hep-ph/9512439].

\bibitem{GarciaBellido:1996qt}
  J.~Garcia-Bellido, A.~D.~Linde and D.~Wands,
  {\it ``Density perturbations and black hole formation in hybrid inflation,''}
  Phys.\ Rev.\  {\bf D 54} (1996) 6040-6058 [astro-ph/9605094].

\bibitem{Copeland:1994vg}
  E.~J.~Copeland, A.~R.~Liddle, D.~H.~Lyth, E.~D.~Stewart and D.~Wands,
  {\it ``False vacuum inflation with Einstein gravity,''}
  Phys.\ Rev.\  {\bf D 49} (1994) 6410-6433 [astro-ph/9401011].

\bibitem{Bugaev:2010bb}
  E.~Bugaev and P.~Klimai,
  {\it ``Constraints on the induced gravitational wave background from primordial black holes,''}
  Phys.\ Rev.\  {\bf D 83} (2011) 083521 [arXiv:1012.4697].
  %%CITATION = PHRVA,D83,083521;%%

\bibitem{Giovannini:2010tk}
  M.~Giovannini,
  {\it ``Secondary graviton spectra and waterfall-like fields,''}
  Phys.\ Rev.\ {\bf D 82} (2010) 083523 [arXiv:1008.1164].
  %%CITATION = PHRVA,D82,083523;%%


\end{thebibliography}
\end{document}